\documentclass{article}
\usepackage{graphicx,amsmath,amssymb,epsfig}

\def\beq{\begin{equation}}
\def\eeq{\end{equation}}

\def\beqn{\begin{eqnarray}}
\def\eeqn{\end{eqnarray}}

\begin{document}
\title{\Large \bf Integrability of scattering amplitudes in $N=4$ SUSY
\footnote{The talk given at the memorial Alexei Zamolodchikov conference,
June 21-23, 2008, Moscow, Russia. }}
\author{\large L.~N. Lipatov
\footnote{Supported by the Maria Curie Award
and grants RFBR 07-02-00902-a, RSGSS 5788.2006.2.}
\bigskip \\
{\it  
St. Petersburg Nuclear Physics Institute, Russia}\\
{\it  
II. Institut f\"{u}r Theoretische Physik, Universit\"{a}t Hamburg, Germany}}

\maketitle

\vspace{-9cm}
\begin{flushright}
{\small DESY--09--020}
\end{flushright}
\vspace{7cm}
\begin{abstract}
\noindent
We argue, that the multi-particle scattering amplitudes in $N=4$
SUSY at large $N_c$ and in the
multi-Regge kinematics for some physical regions have the high
energy behavior appearing from the contribution of the Mandelstam cuts
in the corresponding $t$-channel partial waves. The Mandelstam cuts
correspond to gluon composite states  in the adjoint
representation of the gauge group $SU(N_c)$. The hamiltonian for
these states in the leading logarithmic approximation coincides with the
local hamiltonian of an integrable open spin chain. We construct the
corresponding wave functions using the integrals of motion and the
Baxter-Sklyanin approach.

\end{abstract}

\section{Introduction}

At high energies $s\gg -t$ in QCD the elastic scattering amplitude
for the process $AB\rightarrow A'B'$ in the
leading logarithmic approximation (LLA)
\beq
\alpha _s \,\ln s \sim 1\,,\,\,\alpha _s \ll 1
\eeq
has the Regge form~\cite{BFKL}
\begin{equation}
A_{2\rightarrow 2}=2\,g\delta _{\lambda _{A}\lambda _{A'}}
T_{AA'}^c\frac{s^{1+\omega (t)}}{t}\,g\,T_{BB'}^c
\,\delta _{\lambda _{B}\lambda _{B'}}\,,\,\,t=-\vec{q}^{{2}}\,.
\end{equation}
Here $T^c$ are the generators of the gauge group $SU(N_c)$,
$\lambda _r$ are the particle helicities and
$j(t)=1+\omega (t)$ is the gluon Regge trajectory
for the space-time dimension $D=4-2\epsilon$
\begin{equation}
\omega (-\vec{q}^{2})=-\frac{\alpha_{s} N_c}{(2\pi )^2}\,
(2\pi \mu )^{2\epsilon}\,\int
d^{2-2\epsilon }k
\,\frac{
\vec{q}^{2}}{\vec{k}^{2}(\vec{q}-\vec{k})^{2}}\approx
-\,a\,\left(\ln
\frac{\vec{q}^{2}}{\mu ^2}-\frac{1}{\epsilon}\right)\,.
\end{equation}
In the framework of the dimensional regularization
the parameter $\mu$ is the
renormalization point for the 't Hooft coupling constant and
\begin{equation}
a=\frac{\alpha _{s}\,N_c}{2\pi }\,\left(4\pi e^{-\gamma}\right)^\epsilon
\,,\,\,\gamma =-\psi (1)\,,
\end{equation}
where $\gamma =-\psi (1)$ is the Euler constant and
$\psi (x)=(\ln \Gamma (x))'$.
The gluon trajectory $j(t)$ was calculated also in the next-to-leading
approximation in QCD~\cite{trajQCD} and in the SUSY gauge
models~\cite{trajN4}.

In LLA gluons with momenta $k_r$ (r=1,..,n) are produced in the
multi-Regge kinematics
\begin{equation}
s=(p_A+p_B)^2\gg s_r=(k_r+k_{r-1})^2\gg -t_r=
{\bf q}_r^2\,,\,\,k_r=q_{r+1}-q_r\,,
\label{multReg}
\end{equation}
where the amplitude has the factorized form
\begin{eqnarray}
A_{2\rightarrow 2+n} =
2\,s\,\delta _{\lambda _{A}\lambda _{A'}}
g \, T^{c_1}_{AA'}
\frac{s_1^{\omega (-\vec{q}_1^2)}}{\vec{q}_1^2}gC_{\mu}(q_2,q_1)
e^*_\mu (k_1)T^{d_1}_{c_2c_1}\frac{s_2^{\omega
(-\vec{q}_2^2)}}{\vec{q}_2^2}
...\frac{s_{n+1}^{\omega (-\vec{q}_{n+1}^2)}}{\vec{q}_{n+1}^2}
\,g\,T^{c_{n+1}}_{BB'}\,\delta _{\lambda _{}\lambda _{B'}}\,.
\end{eqnarray}
Here $C_\mu (q_2,q_1)$ is the
effective Reggeon-Reggeon-gluon vertex. In the case when the polarization
vector $e_{\mu}(k_1)$ describes the gluon with a positive
helicity in its c.m. system with the particle $A'$
one can obtain~\cite{effmult}
\begin{equation}
C\equiv C_\mu
(q_2,q_1)\,e^*_{\mu}(k_1)=\sqrt{2}\,\frac{q_2^*q_1}{k_1}\,,
\label{helicityproduction}
\end{equation}
where the complex notation $q=q_x+iq_y$ for the two-dimensional transverse
vectors $\vec{q}$ was used.

The elastic scattering amplitude with vacuum quantum numbers in the
$t$-channel is calculated in terms of
the production amplitude $A_{2\rightarrow 2+n}$ with the
use of the $s$-channel
unitarity~\cite{BFKL}. In this approach the Pomeron appears as a composite
state of two Reggeized gluons. It is convenient to present the gluon
transverse coordinates in the complex form together with their canonically
conjugated momenta~\cite{effmult, int1}
\begin{equation}
\rho
_{k}=x_{k}+iy_{k}\,,\,\,\rho
_{k}^{\ast
}=x_{k}-iy_{k}\,,\,\,p_{k}=i
\frac{\partial }{\partial \rho
_{k}}\,,\,\,p_{k}^{\ast }=
i\frac{\partial }{\partial \rho
_{k}^{\ast }}\,.
\end{equation}
In this case the homogeneous Balitsky-Fadin-Kuraev-Lipatov (BFKL)
equation for the Pomeron wave function can be written as
follows~\cite{BFKL}
\begin{equation}
E\,\Psi
(\vec{\rho}_{1},\vec{\rho}_{2})=
H_{12}\,\Psi (\vec{\rho}_{1},\vec{%
\rho}_{2})\;,\,\,\Delta
=-\frac{\alpha _{s}N_{c}}{2\pi
}\,\min \,E\,,
\end{equation}
where $\Delta$ is the Pomeron intercept entering in the asymptotic
expression
for the total cross-section $\sigma _t\sim s^{\Delta}$.
The BFKL Hamiltonian has a
rather simple operator representation~\cite{int1}
\begin{equation}
H_{12}=\ln
\,|p_{1}p_{2}|^{2}+\frac{1}{p_{1}p_{2}^{\ast
}}(\ln \,|\rho _{12}|^{2})\,p_{1}p_{2}^{\ast
}+\frac{1}{p_{1}^{\ast
}p_{2}}(\ln \,|\rho
_{12}|^{2})\,p_{1}^{\ast
}p_{2}-4\psi (1)
\label{H12}
\end{equation}
with $\rho _{12}=\rho _1-\rho
_2$. The kinetic energy is proportional to the sum of two gluon Regge
trajectories $\omega (-|p_i|^2)$ ($i=1, 2$).
The potential energy $\sim \ln \,|\rho_{12}|^{2}$ is obtained by the
Fourier transformation from the product of
two gluon production vertices $C_\mu$. This Hamiltonian is invariant under
the
M\"{o}bius transformation~\cite{moeb}
\begin{equation}
\rho _{k}\rightarrow
\frac{a\rho _{k}+b}{c\rho
_{k}+d}\,,
\end{equation}
where $a,b,c$ and $d$ are complex parameters. The eigenvalues of the
corresponding Casimir operators are expressed in terms of the conformal
weights
\begin{equation}
m=\frac{1}{2}+i\nu
+\frac{n}{2}\,,\,\,\widetilde{m}=\frac{1}{2}+i\nu
-\frac{n}{2}\,,
\end{equation}
where $\nu$ and $n$ are respectively real and integer numbers
for the principal series of unitary
representations of the M\"{o}bius group $SL(2,C)$.
The eigenvalues of $H_{12}$ depend
on these parameters~\cite{moeb}
\beq
E_{m,\widetilde{m}}=\psi (m)+\psi (1-m)+\psi (\widetilde{m})+
\psi (1-\widetilde{m})-4\psi (1)\,.
\label{PomE}
\eeq
The Pomeron intercept in LLA is positive
\beq
\Delta =4\;\frac{\alpha
_{s}}{\pi }N_{c}\,\ln 2>0
\eeq
and therefore the Froissart
bound $\sigma _t<c\ln ^2s$
for the total cross-section
is violated~\cite{BFKL}.
To restore the broken $s$-channel unitarity one
should take into account the contributions of diagrams corresponding to
the $t$-channel exchange
of an arbitrary number of reggeized gluons in the $t$-channel.
The wave function of the colorless state constructed from $n$ reggeized
gluons can be obtained in LLA as a solution of the
Bartels-Kwiecinski-Praszalowicz (BKP) equation~\cite{BKP}
\beq
E\,\Psi =H^{(0)}\,\Psi \,,\,\,\Delta
=-\frac{\alpha _{s}N_{c}}{4\pi}\,\min \,E\,
\eeq
In the $N_c\rightarrow \infty$ limit the color structure is
simplified and the corresponding Hamiltonian has
the property of the
holomorphic separability~\cite{separ}
\begin{equation}
H^{(0)}=\sum _{k=1}^nH_{k,k+1}=
h^{(0)}+h^{(0)*}\,,\,\,[h^{(0)},h^{(0)*}]=0\,.
\end{equation}
It is a consequence of the similar property
for the pair BFKL hamiltonian $H_{12}$ (\ref{H12}) and the
energy $E_{m,\widetilde{m}}$ (\ref{PomE}).

The holomorphic
Hamiltonian in the multi-color QCD can be written as follows (cf. (\ref{H12}))
\begin{equation}
h^{(0)}=\sum _kh^{(0)}_{k,k+1}\,,\,\,h^{(0)}_{12}=\ln
(p_{1}p_{2})+\frac{1}{p_{1}}\,(\ln
\rho _{12})\,p_{1}+
\frac{1}{p_{2}}\,(\ln \rho
_{12})\,p_{2}-2\psi (1)\,,
\label{pairham}
\end{equation}
where $\psi (x)=(\ln \Gamma (x))'$.
As a result, the wave function $\Psi $ has the holomorphic
factorization~\cite{separ}
\begin{equation}
\Psi =\sum _{r,\widetilde{r}}a_{r,\widetilde{r}}\,\Psi ^{r}(\rho _1,...,\rho _n)
\,\Psi ^{\widetilde{r}}(\rho ^*_1,...,\rho ^*_n)\,,
\end{equation}
which in the case of
two-dimensional
conformal field theories is a consequence of the infinite dimensional Virasoro group.
Moreover, the holomorphic hamiltonian $h^{(0)}$ is invariant under the duality
transformation~\cite{dual}
\begin{equation}
p_i\rightarrow \rho _{i,i+1}\rightarrow p_{i+1}\,,
\end{equation}
combined with its transposition.

Further, there are integrals of motion $q_r$
commuting among themselves and with $h^{(0)}$~\cite{int1, int}:
\begin{equation}
q^{(0)}_{r}=\sum_{k_{1}<k_{2}<...<k_{r}}\rho
_{k_{1}k_{2}}\rho
_{k_{2}k_{3}}...\rho
_{k_{2}k_{3}}...\rho
_{k_{r}k_{1}}\,p_{k_{1}}p_{k_{2}}...
p_{k_{r}}\,,\,\,[q_{r},h]=0\,.
\end{equation}
The integrability of the BFKL dynamics in LLA was established
in Ref. ~\cite{int}. This remarkable property  is
related to the fact that $h$ coincides with the local Hamiltonian of an
integrable Heisenberg spin model \cite{LiFK}. Eigenvalues and
eigenfunctions of this hamiltonian were constructed in Refs. \cite{dVL, DKM}
in the framework of the Baxter-Sklyanin approach~\cite{Sklya}.

In the next-to-leading approximation
the integral kernel for the BFKL equation was
constructed in Refs.~\cite{trajN4,FL}.
In QCD the eigenvalue of the kernel contains
the Kroniker symbols $\delta _{n,0}$ and $\delta _{n,2}$ but in $N=4$ SUSY
it is an analytic function of the conformal spin and
having the property of the maximal
transcendentality~\cite{trajN4,KL}.
This extended supersymmetric theory appears in the framework of
the AdS/CFT correspondence~\cite{Malda, GKP, W}.
It is important, that the one-loop anomalous
dimension for twist-2 operators
in $N=4$ SUSY is proportional
to the expression $\psi (1)-\psi (j-1)$,
which is related to the integrability of evolution
equations for the quasi-partonic operators in
this model~\cite{L4}. The integrability persists
also for some operators in QCD~\cite{BDMB}. The maximal transcendentality
principle suggested in Ref.~\cite{KL} gave a possibility  to extract the
universal anomalous dimension up to three loops in $N=4$
SUSY~\cite{KLV, KLOV} from the corresponding QCD results~\cite{VMV}.
The integrability of the $N=4$ model
was verified also for other operators, large coupling constants
and in higher loops~\cite{MZ, BKSZ, BS}. The asymptotic Bethe ansatz and
integrability allowed to calculate the anomalous dimensions in
four loops~\cite{KLRSV}. The result is in an agreement with the
next-to-leading BFKL predictions
after taking into account the wrapping effects~\cite{BJL}.
The maximal transcendentality was helpful for finding a closed
integral equation for the cusp anomalous
dimension in this model~\cite{ES, BES} with the use of
the 4-loop result~\cite{Bern:2006ew}.

There is another region of investigation, in which
remarkable properties of
the N=4 SUSY are also found. Namely, Bern, Dixon and Smirnov (BDS)
suggested
a simple ansatz for the gluon scattering amplitudes in this model~\cite{BDS}.
This ansatz was verified for the elastic amplitude in the strong coupling
regime using the AdS/CFT correspondence~\cite{AldayMalda}. But the BDS
hypothesis does not agree in this regime with the calculation of the multi-particle
amplitude~\cite{Alday:2007he}. The property of the conformal
invariance of the BDS amplitudes in the momentum space was discussed in
Ref. \cite{DHSS} and the relation with the Wilson loop approach was suggested
in Ref. \cite{Drummond:2007bm} generalizing the results of the strong coupling
calculations of Ref.~\cite{AldayMalda}. The BDS amplitudes $A_n$ for $n\ge 6$
in the multi-Regge kinematics do not have correct analytic properties
compatible with the Steinman relations~\cite{BLV1}. It is a consequence of the
fact, that these amplitudes do not include the Mandelstam cuts~\cite{BLV1}.
This cut contribution was obtained from the BFKL-like
equation for the amplitude with the $t$-channel exchange in the adjoint
representation of the gauge group~\cite{BLV1}. This equation
was solved in LLA
and the two-loop expression for the 6-point
scattering amplitude in the multi-Regge kinematics was derived~\cite{BLV2}.
Recently the two-loop correction was calculated numerically for some values
of external momenta in an agreement with expectations based on the Wilson loop
approach~\cite{twoloop}.

In this paper we demonstrate, that in the multi-color limit for the
production amplitudes the contributions
of the Mandelstam cuts generated by the multi-Reggeon $t$-channel exchange
can be expressed in terms of the solution of the BKP-like equation for the
composite
states of several reggeized gluons in the adjoint representation.
It turns out, that in LLA the corresponding
Hamiltonian coincides with the local Hamiltonian of an
integrable open Heisenberg spin chain. These results partly were presented at the
conferences~\cite{Quarks, Utrecht}.

\section{Mandelstam cuts}

A planar amplitude for the production of two gluons in the multi-Regge
kinematics $s\gg |s_{1}|\sim |s_{2}|\sim |s_{3}|\gg |t_{1}|\sim |t_{2}|
\sim |t_{3}|$ has the multi-Regge form almost in all
physical kinematical regions. But in the physical
region where $s_{1},\,s_{3}<0;\,s>0,\,s_{2}>0$ the amplitude  contains
also the Mandelstam cut~\cite{CutMand} in the angular momentum plane $j_{2}$ of the
crossing channel $t_{2}=-q^{2}$~\cite{BLV1} in the adjoint
representation of the color group. The cut appears as a result of the
exchange of two reggeized gluons with the momenta $p_1=k$ and $p_2=q-k$,
respectively~\cite{BLV2} (see Appendix A for more details).
In the region $s_1,s_3<0;\,s_2>0 $ the integrals over the Sudakov
variables
$\alpha =kp_{A}/p_{A}p_{B}$ and $\beta =kp_{B}/p_{A}p_{B}$ do not vanish
as in other
regions because the integrand contains singularities situated
above and below the corresponding integration contours in an
accordance with the Mandelstam requirements~\cite{CutMand}. These
singularities lead to simultaneous
discontinuities of the amplitude in the invariants $s_{2}$ and $s$.

For the planar amplitude with six external particles
only diagrams with two reggeons in the $t_2$-channel give a non-vanishing
contribution because for
a larger number of reggeons the Mandelstam conditions for singularities
in other Sudakov variables are not fulfilled. However in the case
of a larger
number of external
particles the exchange of several reggeons with momenta $p_l$ gives also
a non-vanishing contribution to the amplitude constructed from planar diagrams.
For the Mandelstam cut resulting from an exchange of $n$ reggeons
one needs at least k=2+2n external particles to have simultaneous singularities in
upper and lower complex semi-planes for the Sudakov parameters
$\alpha '_l, \beta '_l$ of the reggeon momenta $p_l$, as it is demonstrated in
Appendix A.

Let us discuss such composite state of $n$ reggeized gluons in the
adjoint representation (cf. a similar approach for the simple case $n=4$
in Ref. \cite{BLV2}).
One can write the homogeneous BKP equation for
its wave function described by an
amplitude with amputated propagators in the form (see Appendix A)
\begin{equation}
H\Psi =E\Psi \,,\,\,\Delta _n=-\frac{g^2N_c}{16\pi ^2}\,E\,.
\end{equation}
Here $H$ is a redefined hamiltonian obtained after
subtraction of the gluon Regge trajectory $\omega (t)$
containing infrared divergencies.
Namely, the Regge trajectory of the composite state
is~\cite{Quarks, Utrecht}
\begin{equation}
\omega _n(t)=a\left(\frac{1}{\epsilon }-\ln \frac{-t}{\mu ^2}\right)+\Delta _n\,,
\,\,a=\frac{g^2N_c}{8\pi ^2}\left(4\pi e^{-\gamma}\right)^\epsilon \,,
\end{equation}
where $\Delta _n$ is the infrared stable quantity expressed in terms of the energy $E$.

The hamiltonian $H$ in the multi-color limit can be written in the
holomorphically separable form (see Appendix A)
(cf. \cite{BLV2})
\begin{equation}
H=h+h^{\ast }\,,\,\,h=\ln \frac{p_{1}\,p_{n}}{q^{2}}
+
\sum_{r=1}^{n-1}h^t_{r,r+1}\,,\,\,q=\sum _1^np_r\,,
\label{h8hol}
\end{equation}
where the pair hamiltonian $h^t_{r,r+1}$ is transposed to the
corresponding unamputated operator (\ref{pairham})
\begin{equation}
h^t_{r,r+1}=\ln
(p_{r}p_{r+1})+p_{r}\ln (\rho _{r,r+1})\,\frac{1}{p_r}+
p_{r+1}\ln (\rho _{r,r+1})\,\frac{1}{p_{r+1}}+2\gamma \,.
\end{equation}
It is seen from eq. (\ref{h8hol}) that the holomorphic hamiltonian
for the composite state in the adjoint representation differs from
the corresponding expression for the singlet case $h^{(0)}$ (\ref{pairham}) after
its transposition only by the substitution
\beq
h_{n,1}\rightarrow \ln \frac{p_1\,p_n}{q^2}\,,
\eeq
which is related to the fact, that the planar Feynman diagrams have
the topology of a strip and the infrared divergencies in the Regge
trajectories of the particles $1$ and $n$ are not compensated by the
contribution from the pair potential
energy $V_{n,1}$.

It turns out, that the eigenvalues $E$ do not depend on $|q|^2$
due to the scale invariance of $H$,
as it will be demonstrated below. As a result, the $t$-dependence of
$\omega _n(t)$ is the same as in the gluon Regge trajectory.

The transposed holomorphic hamiltonian is related to
the initial hamiltonian by the similarity transformation
\begin{equation}
h^t=\left(\prod _{r=1}^np_r\right)^{-1}\,h\,\left(\prod _{r=1}^np_r\right)\,,
\end{equation}
which leads to the following hermicity property of the total
hamiltonian $H$
\begin{equation}
H^+=\left(\prod _{r=1}^n|p_r|^2\right)^{-1}\,H\,\left(\prod
_{r=1}^n|p_r|^2\right)\,.
\end{equation}
 The last relation is compatible with the normalization
condition
for the wave function
\begin{equation}
||\Psi ||^2=\int \prod _{r=1}^{n-1}d^2p_r \,
\Psi ^*\prod _{s=1}^n|p_s|^{-2}\Psi \,,\,\,\sum _{s=1}^np_s=q\,.
\end{equation}

Using the duality transformation (cf.~\cite{dual})
\begin{equation}
p_1=z_{0,1}\,,\,\,p_r=z_{r-1,r}\,,\,\,q=z_{0,n}\,,\,\,\rho _{r,r+1}=
i\frac{\partial}{\partial z_r}=i\partial _r\,,
\label{dualtr}
\end{equation}
the holomorphic hamiltonian can be rewritten as follows
\begin{equation}
h=\ln \frac{z_{0,1}\,z_{n-1,n}}{z_{0,n}^{2}}
+\sum_{r=1}^{n-1}h^t_{r,r+1}\,,
\label{hol1}
\end{equation}
where
\begin{equation}
h^t_{r,r+1}=2\ln (\partial _r)
+\frac{1}{\partial _r}\,\frac{1}{z_{r-1,r}}+
\frac{1}{\partial _r}\,\frac{1}{z_{r+1,r}}+
\ln (z_{r,r+1}\,z_{r-1,r})+2\gamma \,.
\end{equation}
Here and later we neglect the pure imaginary contribution $2\ln (i)$
because it is cancelled in the total hamiltonian $H$.
Note, that for the colorless composite state and $q=z_{0,n}=0$ the
transformation (\ref{dualtr}) is indeed reduced to the usual duality
substitution of Ref.~\cite{dual}.

To simplify $h$ one can use  the relations~\cite{int1, dual}
\[
\ln \partial =-\ln x +\frac{1}{2}
(\psi (x\partial +1)+\psi (-x\partial ))\,,
\]
\[
\ln (x^2\partial ) =\ln x +\frac{1}{2}
(\psi (x\partial )+\psi (-x\partial +1))\,,
\]
\begin{equation}
\ln \partial =\ln (x^2\partial )-2\ln x +\frac{1}{\partial}\,
\frac{1}{x}\,.
\end{equation}
Then $h^t_{r,r+1}$ can be presented as follows
\begin{equation}
h^t_{r,r+1}=\ln (z_{r,r+1}^2\partial _r)+
\ln (z_{r-1,r}^2\partial _{r})-\ln z_{r,r+1}
-\ln z_{r-1,r}+
2\gamma \,.
\end{equation}

Further, by regrouping its terms we can write the holomorphic
hamiltonian in another form
\begin{equation}
h=-2\ln z_{0,n}
+\ln (z_{0,1}^2\partial _1)+
\ln (z_{n-1,n}^2\partial _{n-1})+2\gamma +
\sum_{r=1}^{n-2}\,h'_{r,r+1}\,,
\label{hamMob}
\end{equation}
where
\[
h'_{r,r+1}=\ln (z_{r,r+1}^2\partial _r)+
\ln (z_{r,r+1}^2\partial _{r+1})-2\ln z_{r,r+1}+
2\gamma
\]
\begin{equation}
=\ln (\partial _r)+\ln (\partial _{r+1})
+\frac{1}{\partial _r}\,\ln z_{r,r+1}\,\partial _r+
\frac{1}{\partial _{r+1}}\,\ln z_{r,r+1}\,\partial _{r+1}
+2\gamma \,.
\label{h2prime}
\end{equation}
The pair hamiltonian $h'_{r,r+1}$ coincides in fact after the
substitution $z_r\rightarrow \rho _r$ with
the corresponding hamiltonian in the coordinate representation
(\ref{pairham}) acting on the wave function with
non-amputated
propagators.

In particular, for $n=2$ one obtains (cf. \cite{BLV2})
\begin{equation}
h=-2\ln z_{0,2}+
\ln (z_{0,1}^2\,\partial _1 )+
\ln (z_{1,2}^2\,\partial_1)
+2\gamma \,.
\end{equation}

It is important, that $h$ (\ref{hamMob}) is invariant under the M\"{o}bius
transformations
\begin{equation}
z_k\rightarrow \frac{az_k+b}{cz_k+d}
\end{equation}
and does not contain the derivatives $\partial _0$ and $\partial _n$.
Therefore we can put
\begin{equation}
z_0=0\,,\,\,z_n=\infty \,,
\end{equation}
which leads to the simplified expression for $h$
\begin{equation}
h\rightarrow h'=
\ln (z_{1}^2\partial _1)+
\ln (\partial _{n-1} )+2\gamma +
\sum_{r=1}^{n-2}\,h'_{r,r+1}\,.
\label{hol2}
\end{equation}
To return to initial variables in the final expression for the wave function
one should perform
the following substitution of $z_k$
\begin{equation}
z_k \rightarrow \frac{z_k-z_0}{z_k-z_n}=
\frac{\sum _{r=1}^kp_r}{q-\sum _{r=1}^kp_r}\,.
\label{substzp}
\end{equation}

According to the above representation (\ref{hol1}) for $h$, its
transposed part $h^t$
can be obtained from $h$ by the similarity transformation which can
be written in terms of $h'$ as follows
\begin{equation}
h^{\prime \,t}=z_1^{-1}\,\left(\prod _{r=1}^{n-2}z_{r,r+1}\right)^{-1}\,
h'\,\,
z_1\,\left(\prod _{r=1}^{n-2}z_{r,r+1}\right)\,,
\end{equation}
which is compatible with the following normalization condition for the
wave function in the full two-dimensional space
\begin{equation}
||\Psi ||^2_1=\int \frac{d^2z_{n-1}}{|z_1|^2}\,\prod _{r=1}^{n-2}
\frac{d^2z _{r}}{|z_{r,r+1}|^2}\,|\Psi |^2\,.
\end{equation}

On the other hand, from the expression (\ref{hol2}) for $h'$
we obtain another relation for $h^{\prime \,t}$
\begin{equation}
h^{\prime \,t}=\left(\prod _{r=1}^{n-1}\partial _r\right)\,
h'\,\left(\prod _{r=1}^{n-1}\partial _r\right)^{-1}\,,
\end{equation}
corresponding to the second normalization condition for $\Psi$
compatible with the hermicity properties of the total hamiltonian
\begin{equation}
||\Psi ||^2_2=\int \prod _{r=1}^{n-1}
d^2z _{r}\,\Psi ^*\prod _{r=1}^{n-1}|\partial _r|^2\,\Psi \,.
\end{equation}

By comparing two above relations between $h'$ and
 $h^{\prime \,t}$ one can conclude (cf. \cite{int}), that the operator
\begin{equation}
A'=z_1\,\prod _{s=1}^{n-2}z_{s,s+1}\,
\prod _{r=1}^{n-1}\partial _r
\label{Aprime}
\end{equation}
commutes with the holomorphic hamiltonian
\begin{equation}
[A',h']=0\,.
\end{equation}

\section{Integrable open spin chain}

Let us verify, that the holomorphic hamiltonian $h'$ (\ref{hol2}) also
commutes with the
differential operator $D(u)$ being the matrix element $T_{22}$ of
the monodromy
matrix (cf.~\cite{int})
\begin{equation}
T(u)=\left(%
\begin{array}{cc}
  A(u) & B(u) \\
  C(u) & D(u) \\
\end{array}%
\right)=L_1(u)L_2(u)...L_{n-1}(u)\,,
\end{equation}
where the $L$-operator is defined by the relation
\begin{equation}
L_r(u)=\left(
\begin{array}{cc}
  u +iz_r\partial _r& i\partial _r \\
  -iz_r^2\partial _r & u-iz_r\partial  _r \\
\end{array}
\right)\,.
\end{equation}

To prove the commutativity of $h'$ and $D(u)$ one can use the
following relation
\begin{equation}
[L_{k}(u)\,L_{k+1}(u), h'_{k,k+1}]=
-i\left(L_{k}(u)-L_{k+1}(u)\right)\,,
\label{LLh}
\end{equation}
valid due to the M\"{o}bius symmetry of the pair hamiltonian
\begin{equation}
[\vec{M}_{k,k+1}, h'_{k,k+1}]=0\,,\,\,\vec{M}_{k,k+1}=
\vec{M}_{k}+\vec{M}_{k+1}
\end{equation}
and the commutation relation (see \cite{dual})
\begin{equation}
[h'_{k,k+1},[\vec{M}^2_{k,k+1},\vec{N}_{k,k+1}]]=
4\vec{N}_{k,k+1}\,,\,\,\vec{N}_{k,k+1}=\vec{M}_{k}-\vec{M}_{k+1}\,.
\label{triline}
\end{equation}
The last relation is a consequence of the fact, that
the operator $\vec{N}_{k,k+1}$ has non-vanishing matrix elements
only between the states $|m_{k,k+1}>$ and $|m_{k,k+1}\pm 1>$ in the
representation, where the Casimir operator of the M\"obius group
is diagonal
\begin{equation}
\vec{M}^2_{k,k+1}|m_{k,k+1}>=m_{k,k+1}(m_{k,k+1}-1)|m_{k,k+1}>\,.
\end{equation}
In this representation the commutation relation (\ref{triline})
is reduced to the recurrent relation for the eigenvalues
$\epsilon (m_{k,k+1})$ of the hamiltonian $h'_{k,k+1}$ (\ref{h2prime})
\begin{equation}
\epsilon (m+1)-\epsilon (m)=2/m \,,
\end{equation}
fulfilled due to the well known representation of $\epsilon (m)$
\begin{equation}
\epsilon (m)=\psi (m)+\psi (1-m)+2\gamma \,.
\end{equation}

Relation (\ref{LLh}) leads to the equality
\begin{equation}
[T(u),\sum_{r=1}^{n-2}h'_{r,r+1}]=
iL_2(u)L_3(u)...L_{n-1}(u)-iL_1(u)L_2(u)...L_{n-2}(u)\,.
\end{equation}
On the other hand, one can easily verify, that
\[
[T_{22}(u),\ln (z_1^2\partial _1)+\ln \partial _{n-1}]=
\left(
0 \,,\,\,1
\right)[T(u),\ln (z_1^2\partial _1)+\ln \partial _{n-1}]
\left(
\begin{array}{cc}
  0 \\
  1 \\
\end{array}
\right)
\]
\begin{equation}
=
-i
\left(
0 \,,\,\, 1
\right)\left(L_2(u)L_3(u)...L_{n-1}(u)-L_1(u)L_2(u)...L_{n-2}(u)
\right)
\left(
\begin{array}{cc}
  0 \\
  1 \\
\end{array}
\right)
\,,
\end{equation}
which proves that the differential operator $D(u)=T_{22}(u)$ is an
integral of motion
\begin{equation}
[D(u),h']=0\,.
\label{commrel}
\end{equation}
Thus, our hamiltonian is the local hamiltonian for an open integrable
Heisenberg spin model with the spins which are generators of the
M\"{o}bius group\footnote{I thank L. D. Faddeev for the fruitful discussion in which
he suggested, that the operator $D(u)$ could be an integral
of motion for this open spin chain.}.

With the use of the following decomposition of the $L$-operators
\begin{equation}
L_r(u)=\left(
\begin{array}{cc}
  u & 0 \\
  0 & u\\
\end{array}
\right)+
\left(
\begin{array}{cc}
  1\\
  -z_r\\
\end{array}
\right)
\left(
 z_r \,,\,\,1
\right)\,i\partial _r
\end{equation}
one can construct the matrix element $T_{22}=D(u)$ in an explicit way
\begin{equation}
D(u)=\sum _{k=0}^{n-1}u^{n-1-k}\,q'_k\,,
\end{equation}
where
\begin{equation}
q'_0=1\,,\,\,q'_{1}=-i\sum _{r=1}^{n-1}z_r\,
\partial _r \,.
\end{equation}
In a general case the integrals of motion $q'_k$ are
given below
\begin{equation}
q'_k=-\sum _{0<r_1<r_2<...<r_{k}<n}z_{r_1}\,
\prod _{s=1}^{k-1}z _{r_s,r_{s+1}}\,
\prod _{t=1}^k i\partial _{r_t}\,.
\end{equation}
In particular, we obtain, that $q'_{n-1}$ is proportional
to the integral of motion $A'$ (\ref{Aprime})
\begin{equation}
q'_{n-1}=-i^{n-1}z_1\prod _{s=1}^{n-2}z_{s,s+1}\,
\prod _{t=1}^{n-1}\partial _t=-i^{n-1}\,A'\,.
\end{equation}

Note, that one can parameterize the monodromy matrix in another form
\begin{equation}
T(u)=\left(
\begin{array}{cc}
  j_0(u) +j_3(u) & j_+(u) \\
  j_-(u) & j_0(u) -j_3(u) \\
\end{array}
\right)\,,\,j_\pm (u)=j_1(u)\pm ij_2(u)\,.
\end{equation}
In this case the Yang-Baxter equations for the currents $j_{\mu }$ have
the Lorentz-invariant representation~\cite{dual}
\begin{equation}
[j_\mu (u),j_\nu (v)]=\frac{\epsilon _{\mu \nu \rho \sigma}}{2(u-v)}
\left(j^\rho (u) j^\sigma (v)-j^\rho (v) j^\sigma (u)\right)\,.
\end{equation}
Here $\epsilon _{\mu \nu \rho \sigma}$ is the antisymmetric tensor
in the four-dimensional Minkowski space and
$\epsilon _{1 2 3 0}=1\,,\,\,g _{\mu \nu} =(1,-1,-1,-1)$.

In particular, we obtain from the Yang-Baxter equations the relation
\begin{equation}
[j_0(u)-j_3(u),j_0(v)-j_3(v)]=[j_0(v),j_3(u)]-[j_0(u),j_3(v)]=0
\end{equation}
and therefore the integrals of motion $q'_k$ are independent
operators and commute each with others
\begin{equation}
[q'_k,q'_l]=0\,.
\end{equation}

\section{Composite states of two and three gluons}

In the case $n=2$ we have only one non-trivial
integral of motion
\begin{equation}
q'_1=-iz_1\,\partial _1 \,.
\end{equation}
Taking into account the normalization
condition for the eigenfunction in the two-dimensional space
\begin{equation}
||\Psi ||^2 =\int \frac{d^2z_1}{|z_1|^2}\,|\Psi |^2\,,
\end{equation}
we find the orthonormalized and complete set of eigenfunctions
\begin{equation}
\Psi _{m,\widetilde{m}}^{(2)}=z_1^{-\frac{1}{2}+m}\,
(z^*_1)^{-\frac{1}{2}+\widetilde{m}}\,,\,\,m=\frac{1+n}{2}+i\nu \,,\,\,
\widetilde{m}=\frac{1+n}{2}-i\nu \,,
\label{wave2}
\end{equation}
satisfying the single-valuedness requirement. Note, that
using the substitution (\ref{substzp}) one can reproduce
the wave functions of two gluon composite states  in the momentum space
(see \cite{BLV2}).

For the case $n=3$ the operator $D(u)$ is given below
\begin{equation}
D_3(u)=u^2-iu(z_1\,\partial _1+z_2 \,\partial _2)+z_1z_{1,2}\,
\partial _1\partial _2 \,.
\end{equation}
With taking into account the normalization condition
\begin{equation}
||\Psi ||^2=\int \frac{d^2z_1\,d^2z_2}{|z_1|^2|z_{1,2}|^2}\,|\Psi |^2\,,
\end{equation}
one can search
the holomorphic eigenfunction of this operator in the form
\begin{equation}
\Psi _m^{(3)}=z_2^{-\frac{1}{2}+m}\,f\left(\frac{z_2}{z_1}\right)\,.
\label{sol3m}
\end{equation}
The function $f(x)$ satisfies the equation
\begin{equation}
\left(x(1-x)\partial ^2+(\frac{1}{2}+m)(1-x)\partial +\lambda \right)
\,f=0\,,\,\,x=\frac{z_2}{z_1}\,,
\end{equation}
where $\lambda $ is the eigenvalue of the operator
$z_1z_{1,2}\partial _1\partial _2$.
Two independent solutions of this equation can be expressed in terms
of the hypergeometric function $F$
\begin{equation}
f_1(x)=F(a_1,a_2;1+a_1+a_2;x)\,,\,\,
f_2(x)=x^{a_1+a_2}\,F(-a_2,-a_1;
1-a_1-a_2;x)\,,
\end{equation}
where the parameters $a_1$ and $a_2$ are obtained from the set of
equations
\begin{equation}
a_1+a_2=-\frac{1}{2}+m\,,\,\,a_1a_2=-\lambda \,.
\label{a1a2}
\end{equation}
The solutions near the point $x=1$ can be also
expressed in terms of hypergeometric functions and are expanded as
follow
\[
\frac{\Gamma (a_1)\,\Gamma (a_2)}{\Gamma (1+a_1+a_2)}
f_1(x)_{|_{x\rightarrow 1}}= \frac{1}{a_1a_2}
-(x-1)
\left(\ln (1-x)-\psi (1)-\psi (2)+\psi (1+a_1)+\psi (1+a_2)\right)
\]
and
\[
\frac{\Gamma (-a_1)\,\Gamma (-a_2)}{\Gamma (1-a_1-a_2)}
f_2(x)_{|_{x\rightarrow 1}}= \frac{1}{a_1a_2}
-(x-1)
\left(\ln (1-x)-\psi (1)-\psi (2)+\psi (1-a_1)+\psi (1-a_2)\right)
\]

Analogously one can find the large-$x$ behavior of the functions
$f_1$ and $f_2$
\[
f_{1}(x)_{|_{x\rightarrow \infty}}=
\frac{\Gamma (a_1+a_2+1)\Gamma (a_2-a_1)}{\Gamma (a_2)\,
\Gamma (1+a_2)}(-x)^{-a_1}
+\frac{\Gamma (a_1+a_2+1)\Gamma (a_1-a_2)}{\Gamma (a_1)\,
\Gamma (1+a_1)}(-x)^{-a_2}\,,
\]
\[
f_{2}(x)_{|_{x\rightarrow \infty}}=
\frac{\Gamma (1-a_1-a_2)\Gamma (a_1-a_2)}{\Gamma (-a_2)\,
\Gamma (1-a_2)}(-x)^{-a_2}
+\frac{\Gamma (1-a_1-a_2)\Gamma (a_2-a_1)}{\Gamma (-a_1)\,
\Gamma (1-a_1)}(-x)^{-a_1}\,.
\]

To construct the wave function $\Psi$
with the property of the single-valuedness in the two-dimensional
subspaces $\vec{z}_1$ and $\vec{x}$
we should write a bilinear combination of the functions $f_i(x)$ and the
corresponding functions in the anti-holomorphic subspace $\widetilde{f}_i(x^*)$
taking into account that in
the second pair of functions one should perform the substitution
\begin{equation}
a_1\rightarrow \widetilde{a}_1\,,\,\,a_2\rightarrow \widetilde{a}_2\,,\,\,
m\rightarrow \widetilde{m}=\frac{1-n}{2}+i\nu \,.
\end{equation}
Due to the single-valuedness of the wave function near $x=0$ we obtain
for it the following expression
\begin{equation}
\Psi =|z_2|^{2i\nu}\,\left(\frac{z_2}{z_2^*}\right)^{\frac{n}{2}}
\Psi
_{m,\widetilde{m}}(\vec{x})\,,\,\,
\Psi
_{m,\widetilde{m}}(\vec{x})=f_1(x)\,\widetilde{f}_1(x^*)+
C\,f_2(x)\,\widetilde{f}_2(x^*)\,,
\label{exprC}
\end{equation}
where the constant $C$ should be fixed from the requirement, that the
analytic continuation of $\Psi $ in the neighborhood of the points
$x=1$ and $x=\infty $ leads also to a single-valued expression.
The condition, that near $x=1$ the terms proportional to
$|x-1|^2\ln (1-x)\ln (1-x^*)$ are absent, gives  the relation
\begin{equation}
\frac{\Gamma (a_1+a_2+1)}{\Gamma (a_1)\,\Gamma (a_2)}\,
\frac{\Gamma (\widetilde{a_1}+\widetilde{a_2}+1)}{\Gamma
(\widetilde{a_1})\,\Gamma (\widetilde{a_2})}+C\,
\frac{\Gamma (1-a_1-a_2)}{\Gamma (-a_1)\,\Gamma (-a_2)}\,
\frac{\Gamma (1-\widetilde{a_1}-\widetilde{a_2})}{\Gamma
(-\widetilde{a_1})\,\Gamma (-\widetilde{a_2})}=0\,.
\end{equation}
Providing, that the constant $C$ is fixed by this equality,
the behavior of the total wave function at $x \rightarrow 1$ is
simplified
\[
\lim _{x\rightarrow 1}\Psi _{m,\widetilde{m}}(\vec{x})\sim
\left(\psi (1+a_1)+\psi (1+a_2)-\psi (-a_1)-\psi (-a_2)\right)
|1-x|^2\ln (1-x^*)
\]
\begin{equation}
+\left(\psi (1+\widetilde{a_1})+\psi (1+\widetilde{a_2})-
\psi (-\widetilde{a_1})
-\psi (-\widetilde{a_2})\right)|1-x|^2\ln (1-x)\,.
\end{equation}

Thus, the single-valuedness condition at $x\rightarrow 1$ leads to
the
additional equation
\begin{equation}
\cot (\pi a_1)+\cot (\pi a_2) =\cot (\pi \widetilde{a_1})+
\cot (\pi \widetilde{a_2})\,.
\end{equation}

A stronger constraint can be obtained from the single-valuedness condition
for $\Psi$ at $x\rightarrow \infty$. Indeed, its consequence for
the bilinear combinations
\[
(-x)^{-a}(-x^*)^{-\widetilde{a_1}}\,,\,\,(-x)^{-a_2}(-x^*)^{-
\widetilde{a_2}}
\]
leads to the relations
\begin{equation}
a_1-\widetilde{a_1}=N_{a_1}\,,\,\,a_2-\widetilde{a_2}=N_{a_2}\,,
\end{equation}
where $N_{a_1},\,N_{a_2}$ are integers. Further,
the absence of the interference terms
\[
(-x)^{-a_1}(-x^*)^{-\widetilde{a_2}}\,,\,\,(-x)^{-a_2}(-x^*)^{-
\widetilde{a_1}}
\]
is fulfilled due to the above relation (\ref{exprC}) for $C$.

One can write the integral representation for the wave function
satisfying the above constraints
\begin{equation}
\Psi \sim z^{a_1+a_2}_2\,(z^*_2)^{\widetilde{a_1}+\widetilde{a_2}}\,
\int \frac{d^2y}{|y|^2}\,
y^{-a_2}(y^*)^{-\widetilde{a_2}}\,\left(\frac{y-1}{y-x}\right)^{a_1}\,
\left(\frac{y^*-1}{y^*-x^*}\right)^{\widetilde{a_1}}\,,\,\,
x=\frac{z_2}{z_1}\,,
\end{equation}
where the integration is performed over the two-dimensional plane $\vec{y}$. Note, that
the integrand has no ambiguity in the points $y=0,1,x$ due to the derived
relations between $a_1,\widetilde{a_1}$
and $a_2,\widetilde{a_2}$. Moreover, the function $\Psi$
near the points $x=0,1, \infty$ can be presented
in terms of the sum of products of above hypergeometric functions.

There is another basis for the holomorphic solutions
\[
\Psi _1(z_1,z_2)=z_1^{a_1}\,z_2^{a_2} \,F(a_1,-a_2,1+a_1-a_2;
\frac{z_1}{z_2})\,,
\]
\begin{equation}
\Psi _2(z_1,z_2)=z_1^{a_2}\,z_2^{a_1} \,
F(a_2,-a_1,1+a_2-a_1; \frac{z_1}{z_2})
\end{equation}
allowing to construct an equivalent representation for the total wave
function $\Psi$.
Note, that these functions can be written in terms of the Mellin-Barnes
integrals
\[
\Psi _1(z_1,z_2) \sim \int _{-i \infty} ^{i\infty}
\frac{\Gamma (a_1+s)\,\Gamma (-a_2+s)\,
\Gamma (-s)}{\Gamma (a_1-a_2+1+s)}\,
(-z_1)^{a_1+s}(-z_2)^{a_2-s}\,d\,s\,,
\]
\begin{equation}
\Psi _2(z_1,z_2) \sim \int _{-i \infty} ^{i\infty}
\frac{\Gamma (a_2+s)\,\Gamma (-a_1+s)\,
\Gamma (-s)}{\Gamma (a_2-a_1+1+s)}\,
(-z_1)^{a_2+s}(-z_2)^{a_1-s}\,d\,s\,.
\end{equation}
Here it is assumed, that the poles of $\Gamma (-s)$ are situated
to the right from the integration contour whereas all other poles
lie to the left of it.

\section{Hamiltonian and integrals of motion}

The holomorphic hamiltonian for composite states of two reggeized gluons
can be written as follows
\begin{equation}
\widetilde{h}=\ln (z_1^2\partial _1)+\ln (\partial _1)+2\gamma
=\psi (z_1\partial _1)+\psi (-z_1 \partial _1)+2\gamma \,.
\end{equation}
Acting by $\widetilde{h}$ on the function $z_1^{\delta}$ we obtain
\begin{equation}
\widetilde{h}z_1^{\delta }=\epsilon (\delta )\,z_1^{\delta}\,,\,\,
\epsilon (\delta )=\psi (\delta )+\psi (-\delta )+2\gamma \,.
\end{equation}
In the case of wave function (\ref{wave2}) satisfying the
single-valuedness and
orthonormality conditions
in the two-dimensional space one derives the following expression for
the total energy~\cite{BLV2}
\begin{equation}
E_{m,\widetilde{m}}=\epsilon _m +\epsilon _{\widetilde{m}}\,,\,\,
\epsilon _m=\psi (-\frac{1}{2}+m)+
\psi (\frac{1}{2}-m)+2\gamma \,.
\end{equation}
Note, that it does not coincide with the corresponding
result (\ref{PomE}) for the Pomeron state.

The holomorphic hamiltonian for composite states of three gluons has
the form
\begin{equation}
h'=\ln (z_1^2\partial _1)+\ln (\partial _2)+\ln (z_{1,2}^2\partial _1)+
\ln (z_{1,2}^2\partial _2)-2\ln  z_{1,2}+4\gamma \,.
\end{equation}
In the region
\begin{equation}
z_1\ll z_2
\end{equation}
it is a sum of two independent pair hamiltonians
\begin{equation}
h'=\psi (z_1\partial _1)+\psi (-z_1 \partial _1)+
\psi (z_2\partial _1)+\psi (-z_2 \partial _1)+4\gamma \,.
\end{equation}
Because the limit $x=z_2/z_1 \rightarrow \infty$ in solution
(\ref{sol3m}) corresponds to this
kinematics,
we obtain
\begin{equation}
\epsilon =\epsilon (a_1)+\epsilon (a_2)\,,
\end{equation}
where $a_1$ and $a_2$ are parameters of the
three-gluon composite state (see (\ref{a1a2})).
The eigenvalues of the integrals of motion are also expressed in terms of
these parameters. Due
to the
normalizability condition these quantities together with the parameters
$\widetilde{a_1},\widetilde{a_2}$ of the wave function in the
anti-holomorphic
space should be chosen as follows
\[
a_1=i\nu _{a_1}+\frac{n_{a_1}}{2}\,,\,\,a_2=
i\nu _{a_2}+\frac{n_{a_2}}{2}\,,
\]
\begin{equation}
\widetilde{a_1}=i\nu _{a_1}-\frac{n_{a_1}}{2}\,,\,\,
\widetilde{a_2}=i\nu _{a_2}-\frac{n_{a_2}}{2}\,,
\label{quanta1a2}
\end{equation}
where $\nu _r$ are real and $n_r$ are integer numbers.

Note, that
\begin{equation}
\nu =\nu _{a_1}+\nu _{a_2}\,,\,\,n=n_{a_1}+n_{a_2}
\,,\,\,a_1a_2=-\lambda \,,
\,\,\widetilde{a_1}\widetilde{a_2}=
-\widetilde{\lambda}\,
\end{equation}
and the eigenvalues of two integrals of motion $q'_k$ can be obtained as
coefficients of the polynomials
\begin{equation}
P_2(u)=(u-ia_1)(u-ia_2)\,,\,\,\widetilde{P}_2(u)=
(u-i\widetilde{a_1})(u-i\widetilde{a_2})\,.
\end{equation}

Generally for the composite state of $n$ reggeized gluons the
situation is similar. Namely, the holomorphic wave function in the
region
\begin{equation}
z_1 \ll z_2 \ll z_3 \ll ...\ll z_{n-1}\,.
\end{equation}
is factorized
\begin{equation}
\Psi _{a_1,a_2,...,a_{n-1}}=\prod _{r=1}^{n-1}z_r^{a_r}\,.
\end{equation}
The energy for this solution is the sum of the particle energies
\begin{equation}
\epsilon =\sum _{r=1}^{n-1}\epsilon (a_r)\,.
\end{equation}

The eigenvalues of integrals of motion $q'_k$ can be expressed in terms of the
coefficients of the polynomial
\begin{equation}
P_n(u)=\prod _{r=1}^{n-1}(u-ia_r)\,.
\end{equation}
Due to the condition of the normalizability the parameters should
have the form
\begin{equation}
a_r=i\nu _r+\frac{n_r}{2}\,,
\end{equation}
where $\nu_r$ is real and $n_r$ is an integer number. The energies and
eigenvalues of the integrals of motion in
the anti-holomorphic space are given by the same expressions
with the corresponding substitution of parameters
\begin{equation}
a_r \rightarrow \widetilde{a}_r=i\nu _r -\frac{n_r}{2}\,.
\end{equation}

The holomorphic wave function satisfies a set of differential equations
following from the eigenvalue equation for the operator $D(u)$
\begin{equation}
D(u)\,\Psi _{a_1,a_2,...,a_{n-1}}=\prod _{r=1}^{n-1}(u-ia_r)\,
\Psi _{a_1,a_2,...,a_{n-1}}\,.
\end{equation}
This equation can be solved with the use of the Taylor expansion
\begin{equation}
\Psi _{a_1,..a_{n-1}}=\prod _{r=1}^{n-1}z_r^{a_r}
\sum _{s_2=0}^\infty  \left(\frac{z_1}{z_2}\right)^{s_2}
...
\sum _{s_{n-1}}^\infty \left(\frac{z_{n-2}}{z_{n-1}}\right)^{s_{n-1}}\,
c(s_2,...,s_{n-1})\,,
\end{equation}
where the coefficients $c(s_2,...,s_{n-1})$ are calculated
in a recurrent way. The recurrent relations obtained from the eigenvalue
equations for different operators $q'_r$ are compatible due to
their commutativity. The obtained solution has the singularities
at $z_{kl}=0$. But, if we consider $(n-1)!$ functions
$\Psi _{a_{i_1},..a_{i_{n-1}}}$ obtained by all
possible
permutations of parameters $a_r$ and multiply them on the corresponding
functions in the anti-holomorphic subspace, it is possible to construct the
wave function having the single-valuedness property in two-dimensional
spaces $\vec{z}_r$
\begin{equation}
\Psi (\vec{z}_1,...,\vec{z}_{n-1})=
\sum _{\{i_1,i_2,...,i_{n-1}\}}
C_{\{i_1,...,i_{n-1}\}}\,\Psi _{a_{i_1},a_{i_2},...,a_{i_{n-1}}}\,
\Psi _{\widetilde{a}_{i_1},\widetilde{a}_{i_2},...,
\widetilde{a}_{i_{n-1}}}\,.
\end{equation}
For this purpose one should adjust the coefficients $C_{\{i_1,...,i_{n-1}\}}$
in an appropriate way presumably without additional constraints on the
parameters $a_r$ and
$\widetilde{a}_r$.
The composite state  of $n-1$ gluons has the following total energy
\begin{equation}
E=\epsilon +\widetilde{\epsilon }\,,\,\,\epsilon =
\sum _{r=1}^{n-1}\epsilon (a_r)\,,\,\,\widetilde{\epsilon}=
\sum _{r=1}^{n-1}\epsilon (\widetilde{a}_r)\,.
\end{equation}

\section{Baxter-Sklyanin approach}

To find a solution of the Yang-Baxter equation for the open spin chain one
can use the Bethe ansatz. For this purpose it is convenient to work
in the transposed representation for the monodromy matrix
\begin{equation}
T^t(u)=\left(%
\begin{array}{cc}
  j^t_0(u)+j^t_3(u) & j^t_+(u) \\
  j^t_-(u) & j^t_0(u)-j^t_3(u) \\
\end{array}%
\right)=L^t_{1}(u)L^t_{2}(u)...L^t_{n-1}(u)\,,
\end{equation}
where the $L$-operator can be chosen as follows
\begin{equation}
L^t_r(u)=\left(
\begin{array}{cc}
  u +i\partial _rz_r& i\partial _r \\
  -i\partial _rz_r^2 & u-i\partial  _r z_r\\
\end{array}
\right)\,.
\end{equation}

The pseudo-vacuum state is defined as a solution of the equation
\begin{equation}
j^t_-(u)\Psi _0=0\,.
\end{equation}
It can be written in the form~\cite{LiFK}
\begin{equation}
\Psi _0=\prod_{r=1}^{n-1}z_r^{-2}\,.
\end{equation}
Note, that the function $|\Psi _0|^2$ does not belong to the principal
series of the unitary representations. As a result, the states constructed
in the framework of the Bethe ansatz by applying  the product
of the operators $j^r_+(u_r)$ to $\Psi _0$
\begin{equation}
\Psi ^t_k =\prod _{r=1}^kj^t_+(u_r)\,\Psi _0
\end{equation}
are non-physical. Nevertheless, these states are eigenfunctions of
the integral of motion
\begin{equation}
D^t(u)\Psi ^t _k=(j^t_0(u)-j^t_3(u))\Psi ^t_k=\Lambda (u)\Psi ^t_k
\end{equation}
providing that
\begin{equation}
\Lambda (u)=(u+i)^{n-1}\prod _{t=1}^k\frac{u-u_t+i}{u-u_t} \equiv
(u+i)^{n-1}\frac{Q(u+i)}{Q(u)}
\end{equation}
is a polynomial, which leads to a quantization condition for
the Bethe roots $u_t$. If we parameterize this polynomial as follows
\begin{equation}
\Lambda (u)=\prod _{l=1}^{n-1}(u-ia_l)\,,
\end{equation}
the above defined Baxter function $Q(u)$ can be calculated
\begin{equation}
Q(u)=\phi (u)\prod _{l=1}^{n-1}\frac {\Gamma (-iu-a_l)}{\Gamma (-iu+1)}\,.
\end{equation}
Here for generality we included the factor $\phi (u)$ which is an
arbitrary periodic function
\begin{equation}
\phi (u)=\phi (u+i)\,.
\end{equation}

In the case of a finite number of the multipliers $j_+^t(u_r)$ in the Bethe
ansatz for the wave function $\Psi _k$ the expression $Q(u)$ is also a
polynomial
\begin{equation}
Q(u)=\prod _{r=1}^{k}(u-u_r)\,.
\end{equation}
For such solutions the parameters $a_l=-k_l-1$ are negative integer numbers
satisfying the condition
\begin{equation}
\sum _{l=1}^{n-1}k_l=k\,.
\end{equation}
The corresponding Baxter functions can be written as follows
\begin{equation}
Q(u)=\prod _{l=1}^{n-1}\prod _{t=1}^{k_l}(u+it)=
\prod _{p=1}^{\max _tk_t}(u+ip)^{r_p}\,,
\end{equation}
where $r_p$ is the number of $k_t$ satisfying the condition
$k_t\geq p$.

As it was mentioned above, the polynomial solutions for $Q(u)$ are non-physical,
because the
corresponding wave functions $\Psi $ do not belong to the principal
series of unitary representations of the M\"{o}bius group. We should find
a set of non-polynomial solutions $Q_{s}(u)$ satisfying this physical
requirement.

According to E. Sklyanin~\cite{Sklya} the correct variables in which the
dynamics of
the Heisenberg spin model is drastically simplified are the zeroes
$\hat{b}_r$ of the
operator $B(u)=j^t_+(u)$ entering in the monodromy matrix
\begin{equation}
B(u)=P_{n-1}\,\prod _{k=1}^{n-2}(u-\hat{b}_r)\,,\,\,
P_{n-1}=i\sum _{r=1}^{n-1}\partial _r \,,
\end{equation}
where the operators $\hat{b}_r$ and $P_{n-1}$
commute each with others
\begin{equation}
[\hat{b}_r,\hat{b_s}]=[\hat{b}_r,P_{n-1}]=0\,.
\end{equation}

It is convenient to pass from the coordinate representation $\vec{z}$ to
the Baxter-Sklyanin representation~\cite{dVL}, in which the currents
$j^t_+(u)$ and $(j^t_+(u))^*$
(together with the operators
$\hat{b}_r, \hat{b}_r^*$ and $P_{n-1}, P_{n-1}^*$)
are diagonal. We denote the eigenvalues of the Sklyanin operators by
$b_r, b_r^*$.
The kernel of the unitary transformation to the Baxter-Sklyanin representation
is known explicitly for the cases $n=2$, $n=3$ and $n=4$~\cite{dVL}.
For general $n$ this integral operator can be presented as a
multi-dimensional integral~\cite{DKM}.

In the Baxter-Sklyanin representation  the wave function in the
holomorphic subspace can be expressed as a
product of the pseudo-vacuum state in this representation
$\Psi _0(P_{n-1},b_1,b_2,...,b_{n-2})$ and the Baxter
functions $Q(u_t)$
\begin{equation}
\Psi ^t(P_{n-1};b_1,...,b_{n-2})=
P _{n-1}^{-\frac{n-1}{2}-m}\prod _{k=1}^{n-2}Q(b_k)\,
\Psi _0(P_{n-1},b_1,...,b_{n-2})\,,
\end{equation}
where the power-like behavior in the variable $P_{n-1}$ is in an
agreement with the normalization condition.

The analogous representation is valid for the total wave function
\begin{equation}
\Psi ^+(\vec{P}_{n-1};\vec{b}_1,...,\vec{b}_{n-2})=
P _{n-1}^{-\frac{n-1}{2}-m}\,(P^* _{n-1})^{-\frac{n-1}{2}-\widetilde{m}}
\prod _{k=1}^{n-2}Q(\vec{b}_r)\,
\Psi _0(\vec{P}_{n-1};\vec{b}_1,...,\vec{b}_{n-2})
\end{equation}
with the use of the generalized
Baxter function $Q(\vec{u})$ being a bilinear combination of the usual
Baxter functions in the holomorphic and anti-holomorphic subspaces
\begin{equation}
Q(\vec{u})=\sum _{s,t}d_{s,t}\,Q_{s}(u)\,Q_t(u^*)\,.
\end{equation}
Here $Q_{s}(u)$ are different solutions of the Baxter equation with
the same eigenvalue $\Lambda (u)$. The coefficients $d_{s,t}$ are chosen
from the requirement, that the function $Q(\vec{u})$ satisfies the
normalization condition everywhere including the points where the
functions $Q_{s}(u)$ and $Q_t(u^*)$ have the poles~\cite{dVL, DKM}.
For the periodic spin chain this condition leads to the
quantization of the eigenvalue of the operator $A(u)+B(u)$ although
a simpler method of quantization is based on the
requirement, that all Baxter functions corresponding to the
same eigenvalue should have the same holomorphic energy~\cite{dVL}. In
the case of the open Heisenberg spin model the situation is simpler
and will be discussed below.

\section{Baxter-Sklyanin representation for two and three gluon states}
Let us consider the composite states constructed
from two and three reggeons in the framework of the Baxter-Sklyanin approach.
In the case $n=2$ we have the following integral
of motion in the transposed space
\begin{equation}
D^t (u)=j_0-j_3=u-i\partial _1z_1
\end{equation}
and its eigenstates in accordance with the Sklyanin approach
are given by the expression
\begin{equation}
\Psi ^t \sim p_1^{-\frac{1}{2}-m}\,z_1^{-2}\sim z_1^{-\frac{3}{2}+m}\,.
\end{equation}
The corresponding transposed hamiltonian is presented below
\begin{equation}
h^t=\ln (\partial _1z_1^2)+\ln \partial _2+2\gamma \,.
\end{equation}
Its eigenvalue calculated on the above eigenfunction $\Psi ^t$ is
\begin{equation}
\epsilon _m =\psi (-\frac{1}{2}+m)+\psi (\frac{1}{2}-m)+2\gamma \,.
\end{equation}

For the states composed from three reggeized gluons the
transposed integral
of motion in the holomorphic subspace is
\begin{equation}
D^t_3(u)=u^2-iu(\partial _1\,z_1+\partial _2\,z_2)+
\partial _1\partial _2 \,z_1z_{1,2}
\end{equation}
and the operator $j_+^t$ is given below
\begin{equation}
j_+^t=iu(\partial _1+\partial _2)-\partial _1\partial _2z_{12}=
i(\partial _1+\partial _2)\,(u-\hat{b}_1)\,,
\end{equation}
where
\begin{equation}
\hat{b}_1=-i\frac{\partial _1\partial _2}{\partial _1+\partial _2}
\,z_{12}\,.
\end{equation}
The operator $j_+^t$ is easily diagonalized after a transition to
the momentum representation, where
\begin{equation}
i\partial _1\,f_{p_1,p_2}=p_1\,f_{p_1,p_2}\,,\,\,
i\partial _2\,f_{p_1,p_2}=p_2\,f_{p_1,p_2}\,.
\end{equation}
In this case the eigenvalue equation for $j_-^t$ has the form
\begin{equation}
\left(u(p_1+p_2)-i\,p_1p_2(\frac{\partial }{\partial p_1}-
\frac{\partial }{\partial p_2})\right)f =(p_1+p_2)(u-b_1)\,f
\,,
\end{equation}
where $b_1$ is the eigenvalue of $\hat{b}_1$.
Its solution  is given below
\begin{equation}
f =\chi (p_1+p_2, b_1)\,\left(\frac{p_1}{p_2}\right)^{-ib_1}\,,
\end{equation}
where $\chi $ is an arbitrary function of $p_1+p_2$ and $b_1$.
The dependence of $\Psi ^t$ from $p_1+p_2$ is fixed by the normalization
condition
\begin{equation}
\Psi ^t\sim (p_1+p_2)^{-a_1-a_2}\,.
\end{equation}

On the other hand, the eigenvalue equation for the integral of motion in
the momentum space can be written
in the form
\begin{equation}
p_1p_2\frac{\partial }{\partial p_1}\,\left(
\frac{\partial }{\partial p_2}-
\frac{\partial }{\partial p_1} \right)\Psi (p_1,p_2)=
a_1a_2 \Psi (p_1,p_2)\,.
\end{equation}
Using the anzatz
\begin{equation}
\Psi (p_1,p_2)=(p_1+p_2)^{-a_1-a_2}\,\eta
(y)\,,\,\,y=\frac{p_2}{p_1}\,,
\end{equation}
we obtain the following equation for the function $\eta(y)$
\begin{equation}
\left(y^2\,\partial ^2+(a_1+a_2+1)\,y\,\partial -a_1a_2\right)\eta (y)=
\left(-y^3\,\partial ^2-2\,y^2\,\partial \right)\eta (y)\,.
\end{equation}
There are two independent solutions of this equation
\[
\eta _1 (y)= \sum _{k=1}^{\infty}
\frac {\Gamma (k-a_1)\,\Gamma (k-a_2)\,(-1)^{k-1}\,y^{-k}}{\Gamma (k+1)\Gamma (k)
\Gamma (1-a_1)\Gamma (1-a_2)}=
\]
\[
\frac{1}{y}\,F(1-a_1,1-a_2,2;-\frac{1}{y})
=\frac{\Gamma (a_1-a_2)\,y^{-a_1}}{\Gamma (1-a_2)\,\Gamma (1+a_1)}\,
F(-a_1,1-a_1,1+a_2-a_1;-y)
\]
\begin{equation}
+\frac{\Gamma (a_2-a_1)\,y^{-a_2}}{\Gamma (1-a_1)\,\Gamma (1+a_2)}\,
F(-a_2,1-a_2,1+a_1-a_2;-y)
\end{equation}
and
\[
\eta _2 (y)= \frac{1}{a_1\,a_2}+\sum _{k=1}^{\infty}
\frac {\Gamma (k-a_1)\,\Gamma (k-a_2)\,(-1)^{k-1}\,y^{-k}}{\Gamma (k+1)\Gamma (k)
\Gamma (1-a_1)\Gamma (1-a_2)}\left(\ln y+c_k(a_1,a_2)\right)
\]
\begin{equation}
= -\frac{\Gamma (-a_1)\,\Gamma (+a_2)}{\Gamma (1+a_2-a_1)}
\,y^{-a_1}\,F(-a_1,1-a_1,1+a_2-a_1; -y)\,,
\end{equation}
where
\begin{equation}
c_k(a_1,a_2)=\psi (k)+\psi (k+1)-\psi (k-a_1)-\psi (1-k+a_2)\,.
\end{equation}
One can construct the bilinear combination of these solutions having
the single-valuedness property at $\vec{y}=\infty$
\begin{equation}
\eta (\vec{y}) \sim \eta _1(y)\,\widetilde{\eta }_2(y^*)+
\eta _2(y)\,\widetilde{\eta }_1(y^*)+\widetilde{C}
\,\eta _1(y)\,\widetilde{\eta} _1(y^*)\,.
\end{equation}
On the other hand let us use the above expression for $\eta _1 (y)$
and $\eta _2 (y)$
expressed in terms of the hypergeometric function regular at $y=0$.
To cancel the interference terms
violating the single-valuedness condition
at $y\rightarrow 0$ in the above bilinear
combination for $\eta (\vec{y})$ we should fix $\widetilde{C}$ as follows
\begin{equation}
\widetilde{C}=-\frac{\sin (a_1\pi )\,
\sin (a_2\pi )}{\pi \sin ((a_1-a_2)\pi ) }
=-\frac{\sin (\widetilde{a}_1\pi )\,\sin (\widetilde{a}_2\pi )}{\pi
\sin ((\widetilde{a}_1-\widetilde{a}_2)\pi ) }\,.
\end{equation}
Finally with the use of the integral representation for the hypergeometric
function the wave function $\Psi ^t $ in the momentum space can be written
as follows
\begin{equation}
\Psi ^t(\vec{p}_1,\vec{p}_2 )=(p_1+p_2)^{-a_1-a_2}
(p_1^*+p_2^*)^{-\widetilde{a}_1-\widetilde{a}_2}
\,\phi (\vec{y})\,,
\end{equation}
where $\phi (y)$ is given below
\begin{equation}
\phi (\vec{y})=\int d^2t\,
\left(\frac{1}{t\,y}+1\right)^{a_1}\,
\left(\frac{1}{t^*\,y^*}+1\right)^{\widetilde{a}_1}\,
(1-t)^{a_2-1}\,(1-t^*)^{\widetilde{a}_2-1}
\end{equation}
and satisfies the single valuedness condition in the $\vec{y}$-space
due to the quantization conditions (\ref{quanta1a2}).

The transition to the Baxter-Sklyanin representation
$(u,\widetilde{u})$ corresponds to the
Mellin-type transformation of $\phi (\vec{y})$
\begin{equation}
\phi (u,\widetilde{u})=\int \frac{d^2y}{|y|^2} \,y^{-iu}\,
(y^*)^{-i\widetilde{u}}
\phi (\vec{y})=\int d^2t \,(1-t)^{a_2-1}\,(1-t^*)^{\widetilde{a}_2-1}
\,\chi (\vec{t})\,,
\end{equation}
where
\begin{equation}
-iu=i\nu _u+\frac{N_u}{2}\,,\,\,
-i\widetilde{u}=i\nu _u-\frac{N_u}{2}\,.
\end{equation}
Here $\nu _u$ is a real number and $N_u=0,\pm 1,\pm 2,...$.
The function $\chi$ is given below
\begin{equation}
\chi (\vec{t})=\int \frac{d^2y}{|y|^2}\,y^{-iu}\,(y^*)^{-i\widetilde{u}}
\,\left(\frac{1}{t\,y}+1\right)^{a_1}\,
\left(\frac{1}{t^*\,y^*}+1\right)^{\widetilde{a}_1}=
t^{iu}(t^*)^{i\widetilde{u}}\,c_1\,.
\end{equation}

The corresponding integrals can be calculated explicitly
\[
c_1=\pi \,\frac{\Gamma (1+\widetilde{a}_1)}{\Gamma (-a_1)}
\,\frac{\Gamma (iu)\,\Gamma (-iu-a_1)}{\Gamma (1-i\widetilde{u})\,
\Gamma (1+i\widetilde{u}+\widetilde{a}_1)}\,,
\]
\[
c_2=\int \frac{d^2t}{|1-t|^2} \,(1-t)^{a_2}\,(1-t^*)^{\widetilde{a}_2}\,
t^{iu}(t^*)^{i\widetilde{u}}
=\pi \,\frac{\Gamma (a_2)}{\Gamma (1-\widetilde{a}_2)}
\,\frac{\Gamma (1+i\widetilde{u})\,\Gamma (-iu-a_2)}{
\Gamma (-iu)\,
\Gamma (1+i\widetilde{u}+\widetilde{a}_2)}\,.
\]
Therefore we obtain for $\phi (u,\widetilde{u})$
the following expression
\begin{equation}
\phi (u,\widetilde{u})=
\frac{\pi ^2\Gamma (1+\widetilde{a}_1)
\Gamma (a_2)}{\Gamma (-a_1)\,\Gamma (1-\widetilde{a}_2)}\,
\frac{\Gamma (iu)\Gamma (1+i\widetilde{u})}{\Gamma (-iu)\,
\Gamma (1-i\widetilde{u})}\,\frac{\Gamma (-iu-a_1)
\,\Gamma (-iu-a_2)}{
\Gamma (1+i\widetilde{u}+\widetilde{a}_1)
\Gamma (1+i\widetilde{u}+\widetilde{a}_2)}\,.
\end{equation}

The inverse transformation corresponds to the
Baxter-Sklyanin representation for the wave function
\begin{equation}
\Psi ^t(\vec{p}_1,\vec{p}_2 )=(p_1+p_2)^{-a_1-a_2}
(p_1^*+p_2^*)^{-\widetilde{a}_1-\widetilde{a}_2}
\,\int d^2u\,
\phi (u,\widetilde{u})\,\left(\frac{p_1}{p_2}\right)^{-iu}\,
\left(\frac{p_1^*}{p_2^*}\right)^{-i\widetilde{u}}
\,,
\end{equation}
where
\begin{equation}
-iu=i\nu _u+\frac{N_u}{2}\,,
\,\,-i\widetilde{u}=i\nu _u-\frac{N_u}{2}\,\,,\,\,\,
\int d^2u\equiv \int _{-\infty}^{\infty} d\nu _u \sum
_{N_u=-\infty}^{\infty} \,.
\end{equation}

One can interpret the wave function $\phi (u,\widetilde{u})$ in
the Baxter-Sklyanin
representation as a product of the pseudo-vacuum state
$u\,\widetilde{u}$ and the total Baxter function
\begin{equation}
\phi (u,\widetilde{u})=u\,\widetilde{u}\,Q(u,\widetilde{u})\,,
\end{equation}
where
\begin{equation}
Q(u,\widetilde{u})\sim
\frac{\Gamma (iu)\Gamma (i\widetilde{u})}{\Gamma (1-iu)\,
\Gamma (1-i\widetilde{u})}\,\frac{\Gamma (-iu-a_1)
\,\Gamma (-iu-a_2)}{
\Gamma (1+i\widetilde{u}+\widetilde{a}_1)
\Gamma (1+i\widetilde{u}+\widetilde{a}_2)}\,.
\end{equation}
This expression for $Q(u,\widetilde{u})$ is symmetric to the substitution
\begin{equation}
(u,a_1,a_2) \leftrightarrow (\widetilde{u},\widetilde{a}_1,\widetilde{a}_2)
\end{equation}
and can be written in the factorized form
\begin{equation}
Q(u,\widetilde{u})\sim Q(u,a_1,a_2)\,Q(\widetilde{u},\widetilde{a}_1,
\widetilde{a}_2)\,,
\end{equation}
where
\begin{equation}
Q(u,a_1,a_2)=
\frac{\Gamma
(-iu-a_1)\Gamma (-iu-a_2)}{\Gamma ^2(1-iu)}\,\Phi (u)\,,
\end{equation}
\begin{equation}
\Phi (u)=
\sqrt{\frac{\sin (\pi (-iu-a_1))\,\sin (\pi (-iu-a_2))}{\sin ^2(-i\pi u)}}
\,.
\end{equation}
The expression  $Q(u,a_1,a_2)$ differs from the Baxter function
in the holomorphic space
\begin{equation}
Q(u)=\frac{\Gamma
(-iu-a_1)\Gamma (-iu-a_2)
}{\Gamma
^2(1-iu)}
\end{equation}
only by the periodic function $\Phi (u)$
and therefore it can be considered also as a
Baxter function. Note, however, that the function $\Phi (u)$ contains
a square root singularity and, as a result, the recurrence
relation for the
function $Q(u,\widetilde{u})$ differs
from the similar relation for $Q(u)$ by a sign in its right hand side
\begin{equation}
Q(u+i,\widetilde{u})=-\frac{(u-ia_1)(u-ia_2)}{(u+i)^2}\,
Q(u,\widetilde{u})\,.
\end{equation}
To overcome this problem we can write $Q(u,\widetilde{u})$
as follows
\begin{equation}
Q(u,\widetilde{u})= Q(u)\,Q(\widetilde{u})\,\Phi (u,\widetilde{u})\,,
\end{equation}
where the function $\Phi$ is given below
\begin{equation}
\Phi (u,\widetilde{u})=
\frac{\sin (\pi (i\widetilde{u}+\widetilde{a}_1))\,
\sin (\pi (i\widetilde{u}+\widetilde{a}_2))}{\sin (i\pi u)\,
\sin (i\pi \widetilde{u})}\,.
\end{equation}
This additional factor $\Phi (u,\widetilde{u})$
can be included in the definition of a new pseudo-vacuum state
\begin{equation}
\Psi_0 =\Phi (u,\widetilde{u})\,u\,\widetilde{u}\,.
\end{equation}
Really this pseudo-vacuum state can be considered as the additional
factor for the wave function in the Baxter-Sklyanin representation
providing correct hermicity properties of
the hamiltonian and integrals of motion in
this representation\footnote{I thank Prof. F. Smirnov for discussions
related to
this important interpretation of the pseudo-vacuum state.
} (see also Ref.~\cite{DKM}). We shall return to this problem in
our future publications.

\section{Conclusion}
In this paper we established, that the gluon production amplitudes in the
planar approximation could have the Mandelstam cut contributions in
the multi-regge kinematics at some physical regions. For the cut corresponding
to the composite states of $n$ reggeized gluons the number of external
particles should be  $k\ge 2+2n$. The wave functions of these
states in the adjoint representation satisfy the BFKL-like equation integrable
in LLA and have the property of the holomorphic factorization.
The corresponding holomorphic hamiltonian coincides with the local hamiltonian for
an integrable open Heisenberg spin model. The Baxter equation for this model
is reduced to a simple recurrent relation and can be solved in terms
of the product of the $\Gamma$-functions. We constructed the wave functions
of composite states of 2 and 3 gluons explicitly.
$$
$$
I thank L. Faddeev, F. Smirnov, M. Staudacher, J. Bartels and A. Sabio Vera for
helpful discussions. 

\appendix

\setcounter{equation}{0}

\renewcommand{\theequation}{A.\arabic{equation}}
\section{Mandelstam cuts in planar diagrams}

Here we discuss an appearance of the Mandelstam cuts \cite{CutMand}
in the crossing channels having adjoint representations of the color group $SU(N_c)$
for the planar
Feynman diagrams in the t'Hooft limit $\alpha \ll 1,\,\alpha _cN_c\sim 1$
and calculate the impact factors corresponding to the
multi-reggeon exchange.
To begin with, let us consider the elastic amplitude $A(s,t)$
for the gluon-gluon scattering in the Regge kinematics. It is well known,
that in the leading logarithmic approximation the corresponding
$t$-channel partial wave contains only one reggeized gluon pole. The contribution from
the Pomeron
exchange with color singlet quantum numbers
is suppressed at large $N_c$. The BFKL Pomeron appears as
a composite state of two reggeized gluons and corresponds to the
Mandelstam cut in the $j$-plane of the crossing channel. For the elastic
amplitude the cuts in the adjoint representation appear in non-planar diagrams and
are also suppressed at large $N_c$. Indeed,
according to S. Mandelstam these contributions should have the
following form
\begin{equation}
A(s,t)\sim \int \frac{d^2k_\perp}{(2\pi )^2 i\,s} \,
\frac{(-s)^{j(-\vec{k}^2)}}{\vec{k}^2}\,
\frac{(-s)^{j(-(\vec{q}-\vec{k})^2)}}{(\vec{q} -\vec{k} )^2}
\,\Phi _1(k_\perp ,q)\,\Phi _2(k_\perp ,q)\,,
\end{equation}
where $j(t)$ are the Regge trajectories. The impact factors
$\Phi _r$ are the integrals from the particle-reggeon scattering
amplitudes $f_r$ (including the reggeon residues) over the invariants $s_r$
in the direct channel
\begin{equation}
\Phi _r(k_\perp ,q_\perp )=\int _L \frac{ds_r}{2\pi i} f_r(p_r,k,q)\,,
\,\,s_1=(p_A-k)^2\,,\,\,s _2=(p_B+k)^2
\,.
\end{equation}
Here the integration contour $L$ goes along the real axis
above the right
singularities of $f_r$ and below left ones  according
to the Feynman prescription. Only when the amplitude $f_r$
is constructed
from the diagrams having both these singularities simultaneously the
result of the integration is non-zero because in an opposite case we
can shift the contour $L$ from the real axis to infinity with a vanishing result.
The Mandelstam
cuts are absent
also for the planar amplitude with five external particles.

However, in the case of the six point amplitude there are planar
diagrams in which the Mandelstam cuts are present. Let us denote the
momenta of initial gluons by $p_A,p_B$ and the momenta of final
particles by $p_{A'},k_1,k_2,p_{B'}$ in an accordance with the order
of multiplication of the corresponding color matrices $T_r$. Then
this cut appears in the physical region, where
\begin{equation}
s=(p_A+p_B)^2>0\,,\,\,s_1=(p_{A'}+k_1)^2<0\,,\,\,s_2=(k_1+k_2)^2>0\,,
\,\,s_3=(k_2+p_{B'})^2<0\,.
\end{equation}
This region corresponds to the transition of four particles with their momenta
$p_A,-k_1,-k_2$ and $p_B$ to the two particles with the momenta $p_{A'}$
and $p_{B'}$.
In the multi-Regge
kinematics, where the corresponding Sudakov parameters are strongly ordered
$1\gg -\beta _1 \gg -\beta _2\,,\,\,-\alpha _1
\ll -\alpha _2 \ll 1$,
the integrands in the
impact factors $\Phi _r$
\begin{equation}
\Phi _1(\vec{k} ,\vec{k}_1, \vec{q}_2, )=\int _L \frac{s\,d\alpha }{2\pi i}\,
f_1(p_A,k,k_1,q_2)\,,\,\,
\Phi _2(\vec{k} ,\vec{k}_2, \vec{q}_2, )=\int _L \frac{s\,d\beta}{2\pi i}  \,
f_2(p_B,k,k_2,q_2)
\end{equation}
in the simplest case have only the poles in the
integration variables
$\alpha  \approx 2kp_A/s$ and $\beta \approx 2kp_B/s$.
\[
f_1=  \frac{1}{(p_A-k)^2+i\epsilon}\,
\frac{1}{(k_1+q_2-k)^2+i\epsilon}
=\frac{1}{-s\alpha -\vec{k}^2+i\epsilon}\,
\frac{1}{-s\alpha \beta _1-(\vec{k}_1+\vec{q}_2-\vec{k})^2+i\epsilon}\,,
\]
\begin{equation}
f_2= \frac{1}{(p_B+k)^2+i\epsilon}\,
\frac{1}{(k_2-q_2+k)^2+i\epsilon}=
 \frac{1}{s\beta -\vec{k}^2+i\epsilon}\,
\frac{1}{s\alpha _2\beta -(\vec{k}_2-\vec{q}_2+\vec{k})^2+i\epsilon}\,.
\end{equation}
These poles are situated
above and below the integration
contours $L$ due to the inequalities $\beta _1<0,\,\alpha _2<0$ valid in the
considered kinematical region where $s_1<0,\,s_3<0$.  Therefore the
integrals are non-zero and
can be calculated by residues
\begin{equation}
\Phi _1(\vec{k} ,\vec{k}_1, \vec{q}_2)=
\frac{1}{(\vec{k}_1+\vec{q}_2-\vec{k})^2}\,,\,\,
\Phi _2(\vec{k} ,\vec{k}_2, \vec{q}_2)=
\frac{1}{(\vec{k}_2-\vec{q}_2+\vec{k})^2}\,.
\end{equation}

In the case of production of two gluons with the same helicity  at the
multi-Regge kinematics in the physical region where $s_1<0,\,s_2>0,\,s_3<0$
the amplitude is proportional to the Born expression
\begin{equation}
A_{2\rightarrow 4}=f_{2\rightarrow 4}\,2s
\,g\,T_{A'A}^{c_1}\,\frac{1}{|q_1|^2}\,g\,C(q_2,q_1)\,
T_{c_2c_1}^{d_1}\,\frac{1}{|q_2|^2}\,g\,C(q_3,q_2)\,T_{c_3c_2}^{d_2}\,
\frac{1}{|q_3|^2}\,g\,T_{B'B}^{c_3}\,,
\end{equation}
where the Reggeon-Reggeon gluon vertex $C$ is given above
(see (\ref{helicityproduction})).
In the lowest order approximation the
corresponding proportionality factor $f_{LO}$ for the Mandelstam cut contribution
in the $t_2$-channel contains some additional multipliers from the effective
vertices $C$ in comparison with the
above result (see \cite{effmult, BLV2})
\begin{equation}
f_{LO}=i\,\frac{g^2\,N_c}{4\pi}\,
\int \frac{\mu ^{2\epsilon}d^{2-2\epsilon }k}{(2\pi)^{1-2\epsilon}}\,
\frac{(-s_2)^{j(-\vec{k}^2)-1}}{|k|^2}\,
\frac{(-s_2)^{j(-(\vec{q}_2-\vec{k})^2)-1}}{|q_2-k|^2}\,\widetilde{\Phi}_1\,
\widetilde{\Phi}_2\,,
\end{equation}
where
\begin{equation}
\widetilde{\Phi} _1=q_1^*(k_1+q_2-k)\,(q_2^*-k^*)\,
\Phi _1(\vec{k} ,\vec{k}_1, \vec{q}_2)=\frac{q_1^*(q_2^*-k^*)}{k^*_1+q^*_2-k^*}
\,,
\end{equation}
\begin{equation}
\widetilde{\Phi} _2=q_3
(q_2^*-k^*_2-k^*)\,(q_2-k)\,\Phi _2(\vec{k} ,\vec{k}_2, \vec{q}_2)
=\frac{q_3(q_2-k)}{q_2-k_2-k}
\,.
\end{equation}
In the weak coupling limit $j=1$ and at $\epsilon \rightarrow 0$ the
amplitude $f_{LO}$ is
\begin{equation}
\lim _{j\rightarrow 1}f_{LO}=\pi i\,a\,
\left(\ln \frac{\vec{q}_1^2\vec{q}_2^2}{(\vec{k}_1+\vec{k}_2)^2\mu ^2}-
\frac{1}{\epsilon}\right)\,,\,\,a=\frac{g^2N_c}{8\pi ^2}\,
\left(4\pi \,e^{-\gamma}\right)^\epsilon \,,
\end{equation}
which coincides in this limit with the logarithm of the factor
$C$ introduced in Ref.~\cite{BLV1}. This factor violates the Regge
factorization
of the BDS amplitude in the considered kinematical region due
to the presence of the Mandelstam cut~\cite{BLV1}.

One can take into account the gluon reggeization in the channels
$t_1$ and $t_3$ using the following substitution in the above expressions
\begin{equation}
\frac{1}{-s\alpha -\vec{k}^2+i\epsilon }\rightarrow
-(s\alpha -i\epsilon )^{j(t_1)-2}\,,\,\,
\frac{1}{s\beta -\vec{k}^2+i\epsilon }\rightarrow
-(-s\beta -i\epsilon )^{j(t_3)-2}\,.
\end{equation}
It would lead to the multiplication of the integrand with the
real factor
\begin{equation}
R=\left(\frac{-(\vec{k}_1+\vec{q}_2-\vec{k})^2}{\beta _1}
\right)^{j(t_1)-1}\,
\left(\frac{-(\vec{k}_2-\vec{q}_2+\vec{k})^2}{\alpha _2}
\right)^{j(t_3)-1}\approx \left(-s_1
\right)^{j(t_1)-1}\,
\left(-s_3
\right)^{j(t_3)-1}\,.
\end{equation}
We can include also the diagrams with the reggeized
gluon scattering in the crossing channel.
It leads to
the following expression for the Mandelstam contribution in LLA
\begin{equation}
f_{LLA}^{Mand}=i\,R\,\frac{g^2\,N_c}{4\pi}\,
\int \frac{\mu ^{2\epsilon}d^{2-2\epsilon }k}{(2\pi)^{1-2\epsilon}}\,
\frac{\mu ^{2\epsilon}d^{2-2\epsilon }k'}{(2\pi)^{1-2\epsilon}}
\frac{1}{|k|^2}\,
\frac{1}{|q_2-k|^2}\,G(\vec{k},\vec{k}',\vec{q}_2;\ln (-s_2))
\,\widetilde{\Phi}_1\,
\widetilde{\Phi}_2\,,
\end{equation}
where $G$ is the Green function satisfying the BFKL-like
equation for the octet quantum numbers in $t_2$-channel
\[
\frac{\partial}{\partial \ln s_2}\,
G(\vec{k},\vec{k}',\vec{q}_2;\ln (-s_2))=
K\,G(\vec{k},\vec{k}',\vec{q}_2;\ln (-s_2))\,,
\]
\begin{equation}
G(\vec{k},\vec{k}',\vec{q}_2;0)=
\frac{(2\pi )^{1-2\epsilon}}{\mu ^{2\epsilon}}\,
\delta ^{2-2\epsilon}(k-k')\,.
\end{equation}
Here the operator $K$ in LLA can be expressed in terms of
the Hamiltonian $H$ which does not contain infrared
divergencies
\[
K=\omega (t_2) -\frac{g^2N_c}{16 \pi ^2}\,H\,,\,\,
\omega (t)=a\left(\frac{1}{\epsilon}-\ln \frac{-t}{\mu _2}\right)\,,
\]
\begin{equation}
H=2\ln \frac{|p_1|^2|p_2|^2}{|q_2|^2}+
p_1p_2^*\,\ln |\rho _{12}|^2\,\frac{1}{p_1p_2^*}+
p_1^*p_2\,\ln |\rho _{12}|^2\,\frac{1}{p_1^*p_2}\,,
\end{equation}
where $p_1=k,\,p_2=q-k$.
$$
$$

Let us consider now the Mandelstam cuts constructed from
several reggeons. The non-vanishing contribution from
the exchange of $r+1$ reggeons appears in the planar diagrams only if the
number of the external lines is $n\ge 2r+4$. For the inelastic
transition $2\rightarrow 2+2r$ with the initial and final momenta
$p_A,p_B$ and $p_{A'},k_1,k_2,...,k_{2r},p_{B'}$, respectively, the cut
exists in the crossing channel with the momentum
\begin{equation}
q=p_A-p_{A'}-\sum _{l=1}^rk_l=p_{B'}-p_{B}+\sum _{l=r+1}^{2r}k_l=
\sum _{l=1}^{r+1}q'_l \,,
\end{equation}
where $q'_l$ are momenta of reggeons forming the composite state.
The corresponding amplitude has the form
\begin{equation}
A(p_A,p_{A'},k_1,...,k_{2r},p_{B'},p_B)\sim
\int \prod _{t=1}^r\frac{d^2q'_t}{2\pi \,s}
\,\prod_{l=1}^{r+1}\frac{(-s)^{j(-\vec{q'}_l^2)}}{|q'_l|^2}\Phi _1
(\vec{q}'_1,...,\vec{q}'_{r+1})\Phi _2 (\vec{q}'_1,...,\vec{q}'_{r+1})\,.
\end{equation}
The impact factors $\Phi _{1,2}$ are given in terms of the integrals over
the Sudakov parameters $\alpha '_l=2q'_lp_A/s,\,\beta' _l=2q'_lp_B/s$ from
the reggeon-particle scattering amplitudes $f_{1,2}$
\begin{equation}
\Phi _1=\prod _{l=1}^{r-1}\int _L\frac{s\,d\alpha '_l}{2\pi i}\,f_1\,,\,\,
\Phi _2=\prod _{l=1}^{r-1}\int _L\frac{s\,d\beta '_l}{2\pi i}\,f_2\,.
\end{equation}

In QCD the tree expressions for $f_{1,2}$ appearing
in the planar diagrams are given below
\[
f_1=I_1\,\frac{1}{(p_A-q'_1)^2}\frac{1}{(p_A-k_0-q'_1)^2}...
\frac{1}{(p_A-\sum _{l=1}^{r}q'_l-\sum _{l=0}^{r-2}k_l)^2}
\frac{1}{(p_A-\sum _{l=1}^{r}q'_l-\sum _{l=0}^{r-1}k_l)^2}\,,
\]
\[
f_2=I_2\,\frac{1}{(p_B+q'_1)^2}\frac{1}{(p_B-k_{2r+1}+q'_1)^2}...
\frac{1}{(p_B+\sum _{l=1}^{r}q'_l-\sum _{l=r+3}^{2r+1}k_l)^2}
\frac{1}{(p_B+\sum _{l=1}^{r}q'_l-\sum _{l=r+2}^{2r+1}k_l)^2}\,,
\]
where $k_0=p_{A'},\,k_{2r+1}=p_{B'}$. The additional factors $I_{1,2}$
contain effective reggeon vertices for the production and scattering of the
gluons with the same helicity. They can be written
in the multi-Regge kinematics (\ref{multReg}) as follows
(cf. \cite{effmult})
\[
I_1=\prod _{l=1}^r
\frac{q^{\prime *}_{l+1}(Q-\sum _{t=1}^lq^{\prime }_t
-\sum _{t=1}^{l-1}k_t)}{(Q^*-\sum _{t=1}^{l+1}q^{\prime *}_t
-\sum _{t=1}^{l-1}k_t^*)}\,\prod _{l=1}^r\beta _r\,,
\]
\begin{equation}
I_2=\prod _{l=1}^r
\frac{q^{\prime }_{l+1}(\widetilde{Q}^*+\sum _{t=1}^lq^{\prime *}_t
-\sum _{t=1}^{l-1}k^*_{2r-t+1})}{(\widetilde{Q}+\sum _{t=1}^{l+1}q^{\prime }_t
-\sum _{t=1}^{l-1}k_{2r-t+1})}\,\prod _{l=1}^r\alpha _r\,,
\end{equation}
where $Q=p_A-p_{A'},\,\widetilde{Q}=p_B-p_{B'}$ and the Sudakov variables
of the produced particles
$\alpha _l=2k_lp_A/s,\,\beta _l=2k_lp_B/s$
are strongly ordered
\begin{equation}
1\gg |\beta _1| \gg |\beta _2|...\gg |\beta _{2k}|\,,\,\,|\alpha _1|\ll
|\alpha _2|\ll ...
|\alpha _{2k}|\ll 1\,.
\end{equation}
In these variables the functions
$f_{1,2}$ are given below
\[
f_1=I_1\,\frac{1}{-s\alpha '_1+i\epsilon}
\frac{1}{-s\beta _1\alpha '_1-|Q-q'_1|^2+i\epsilon}
\frac{1}{-s\beta _1\alpha '_2+i\epsilon}
\frac{1}{-s\beta _2\alpha '_2-|Q-q'_1-q'_2-k_1|^2+i\epsilon}
...\,,
\]
\[
f_2=I_2\,\frac{1}{s\beta '_1+i\epsilon}
\frac{1}{s\alpha _{2r}\beta '_1-|\widetilde{Q}+q'_1|^2+i\epsilon}
\frac{1}{s\alpha _{2r}\beta '_2+i\epsilon}
\frac{1}{s\alpha _{2r-1}\beta '_2-|\widetilde{Q}+q'_1+q'_2-k_{2r}|^2+i\epsilon}
...\,,
\]
where we took into account,
that in the essential region of integration
\begin{equation}
\alpha '_l\sim \frac{|Q|^2}{s\beta _l}\,,\,\,
\beta '_l\sim \frac{|\widetilde{Q}|^2}{s\alpha _{2r-l+1}}\,.
\end{equation}
In the physical
region, where the signs of the Sudakov parameters of momenta $k_l$ alternate with
the index $l$
\begin{equation}
\beta _1,\,\alpha _{2r}<0\,;\,\,\beta _2,\,\alpha _{2r-1}>0
\,;\,\,\beta _3,\,\alpha _{2r-2}<0\,;...\,,
\end{equation}
which is equivalent to the following constraints on the invariants
\begin{equation}
s_1<0,s_2<0,...,s_r<0,s_{r+1}>0,s_{r+2}<0,s_{r+3}<0,...,s_{2r+1}<0,s>0\,,
\end{equation}
the integrands in expressions for $\Phi _{1,2}$ contain poles
above and below the
integration contours $L$ over
all variables $\alpha' _l,\,\beta '_l$. Therefore $\Phi _{1,2}$
are non-zero and
can be calculated by taking residues from
the poles in $f_{1,2}$
\begin{equation}
\Phi _1(\vec{q}'_1,...,\vec{q}'_{r+1})=\prod _{l=1}^r
\frac{q^{\prime *}_{l+1}}{(Q^*-\sum _{s=1}^lq^{\prime *}_s-\sum _{s=1}^{l-1}k^*_s)
\,(Q^*-\sum _{t=1}^{l+1}q^{\prime *}_t
-\sum _{t=1}^{l-1}k_t^*)}
\end{equation}
\begin{equation}
\Phi _2(\vec{q}'_1,...,\vec{q}'_{r+1})=
\prod _{l=1}^r\frac{q'_{l+1}}{(\widetilde{Q}+\sum _{s=1}^lq'_s-
\sum _{s=1}^{l-1}k_{2r -s+1})\,(\widetilde{Q}+\sum _{t=1}^{l+1}q^{\prime }_t
-\sum _{t=1}^{l-1}k_{2r-t+1})}\,.
\end{equation}

In the case of the production of $2r$ gluons with the same helicity
the amplitude in $N=4$ SUSY is proportional to the Born expression
containing the
effective Reggeon-Reggeon-gluon vertices $C$ (\ref{helicityproduction}).
The proportionality factor  $f_{2\rightarrow 2+2r}$ for the Mandelstam cut
constructed from $r+1$ reggeized gluons
can be written as follow
\begin{equation}
f_{LO}^{2\rightarrow 2+2r}=\left(i\,\frac{g^2\,N_c}{4\pi}\right)^{r}\,
Q^*\widetilde{Q}\int \prod _{l=1}^r\frac{\mu ^{2\epsilon}d^{2-2\epsilon }
q'_l}{(2\pi)^{1-2\epsilon}}\,\prod _{l=1}^{r+1}
\frac{(-s_{r+1})^{j(-|q'_l|^2)-1}}{|q'_l|^2}\,\Phi _1\,
\Phi_2\,\prod _{t=1}^rk^*_tk_{2r-t}\,.
\end{equation}

In the leading logarithmic approximation the proportionality factor has the
form
\[
f_{LLA}^{2\rightarrow 2+2r}
\]
\begin{equation}
=\left(i\,\frac{g^2\,N_c}{4\pi}\right)^{r}\,
Q^*\widetilde{Q}\int \prod _{l=1}^r\frac{\mu ^{2\epsilon}d^{2-2\epsilon }
p_l}{(2\pi)^{1-2\epsilon}}\,\frac{\mu ^{2\epsilon}d^{2-2\epsilon }
p'_l}{(2\pi)^{1-2\epsilon}}\,\prod _{l=1}^{r+1}
\frac{1}{|p_l|^2}\,G(p,p';s_{r+1})\Phi _1\,
\Phi_2\,\prod _{t=1}^rk^*_tk_{2r-t}\,,
\end{equation}
where we introduce the new notation $p_l$ for the reggeon momenta $q'_l$. The
Green function satisfies the equation
\begin{equation}
\frac{\partial}{\partial \ln s_{r+1}}\,
G(\vec{p},\vec{p}';s_{r+1})=
K\,G(\vec{p},\vec{p}';s_{r+1})\,,\,\,
G(\vec{p},\vec{p}';0)=
\prod _{l=1}^r\frac{(2\pi )^{1-2\epsilon}}{\mu ^{2\epsilon}}\,
\delta ^{2-2\epsilon}(p_{l}-p'_{l})\,.
\end{equation}
Here the operator $K$ in LLA can be expressed in terms of
the Hamiltonian $H$ which does not contain infrared
divergencies
\[
K=\omega (t) -\frac{g^2N_c}{16 \pi ^2}\,H\,,\,\,
\omega (t)=a\left(\frac{1}{\epsilon}-\ln \frac{-t}{\mu _2}\right)\,,
\,\,t=-|q|^2\,,
\]
\begin{equation}
H=\ln \frac{|p_1|^2|p_{r+1}|^2}{|q|^4}+\sum _{l=1}^{r}H_{l,l+1}\,,
\end{equation}
where
\begin{equation}
H_{l,l+1}=\ln |p_l|^2+\ln |p_{l+1}|^2+
p_l\,p_{l+1}^*\,\ln |\rho _{l,l+1}|^2\,\frac{1}{p_l\,p_{l+1}^*}+
p_l^*\,p_{l+1}\,\ln |\rho _{l,l+1}|^2\,\frac{1}{p_l^*\,p_{l+1}}\,.
\end{equation}
Note, that the above hamiltonian has the property of the holomorphic
separability
\begin{equation}
H=h+h^*\,,\,\,h=\ln \frac{p_1\,p_{r+1}}{q^2}+
\sum _{l=1}^{r}h_{l,l+1}\,,
\end{equation}
where
\begin{equation}
h_{l,l+1}=\ln p_l+\ln p_{l+1}+
p_l\,\ln \rho _{l,l+1}\,\frac{1}{p_l}+
p_{l+1}\,\ln \rho _{l,l+1}\,\frac{1}{p_{l+1}}\,.
\end{equation}
One can take into account also the enhanced contributions in the impact
factors leading to the Regge-type dependence of the
amplitude from other invariants $s_i$ ($i\ne r+1$).

\end{document}